\documentclass[10pt,pra,twocolumn,groupaddress,showpacs,floatfix]{revtex4}%
\usepackage{graphics}
\usepackage{amsmath}
\usepackage{graphicx}
\usepackage{amsfonts}
\usepackage{amssymb}%
\begin{document}
\title{Continuous-variable quantum cryptography with an untrusted relay: \\Detailed security analysis of the symmetric configuration}
\date{\today }

\begin{abstract}
We consider the continuous-variable protocol of Pirandola \textit{et al}.
[Nature Photonics \textbf{9}, 397--402 (2015), see also arXiv.1312.4104] where
the secret-key is established by the measurement of an untrusted relay. In
this network protocol, two authorized parties are connected to an untrusted
relay by insecure quantum links. Secret correlations are generated by a
continuous-variable Bell detection\ performed on incoming coherent states. In
the present work we provide a detailed study of the symmetric configuration,
where the relay is midway between the parties. We analyze symmetric
eavesdropping strategies against the quantum links explicitly showing that, at
fixed transmissivity and thermal noise, two-mode coherent attacks are optimal,
manifestly outperforming one-mode collective attacks based on independent
entangling cloners. Such an advantage is shown both in terms of security
threshold and secret-key rate.

\end{abstract}

\author{Carlo Ottaviani}
\email{Carlo.Ottaviani@york.ac.uk}
\author{Gaetana Spedalieri}
\author{Samuel L. Braunstein}
\author{Stefano Pirandola}
\email{Stefano.Pirandola@york.ac.uk}
\affiliation{Department of Computer Science, University of York, York YO10 5GH, United Kingdom}
\maketitle




\section{Introduction}

Rather than layers of physical security or obscure coding in the distribution
of a classical key \cite{ClassCRY}, quantum cryptography claims to rely on the
fundamental laws of quantum physics to provide secure quantum key distribution
(QKD) \cite{Gisin}. In the simplest scheme, information is encoded on
non-orthogonal states which are transmitted from Alice to Bob via a quantum
channel. After classical procedures of error correction and privacy
amplification, Alice and Bob are then able to distill shared secret bits
\cite{Gisin}. Any attempt by Eve at gleaning information inevitably introduces
noise in the quantum channel; by quantifying this noise, Alice and Bob
determine how much distillation must be applied to make Eve's information
negligible or if there is too much noise (breaching the \textit{security
threshold}) they abort the protocol.

Over the last several decades, QKD has been implemented many times, more
recently even in simple network configurations \cite{SECOQC,SECOQC2,Tokyo1}.
By utilizing an iterated point-to-point strategy, a secret key may be shared
with legitimate users on more distant network nodes. This strategy explicitly
requires secure and trusted intermediate relay stations. However, modern
communication protocols (e.g., TCP/IP) rely on a more advanced `end-to-end
principle'~\cite{endtoend,Baran} with simple, unreliable and untrusted relays;
any layers of security are provided by the two end-parties.

A first step in this direction has been done by Ref.~\cite{SidePRL}, which
extended QKD\ to the mediation of an untrusted relay, void of quantum sources
(e.g., entanglement) and performing a quantum measurement which generates
secret correlations in remote stations. This idea of a measurement-based
untrusted relay (also known as `measurement-device independence') has been
experimentally implemented with discrete variables~\cite{EXP1,EXP2}.

More recently, the use of measurement-based untrusted relays has been
introduced in continuous-variable QKD~\cite{RELAY} with the aim of exploiting
the advantages of bosonic systems~\cite{BraREV2,RMP} in terms of cheap quantum
sources and highly-efficient detectors. Indeed this protocol is able to
achieve secret-key rates orders of magnitude higher than any protocol based on
discrete-variable systems. This is possible thanks to the use of the cheapest
possible quantum sources, i.e., coherent states, combined with simple linear
optics and homodyne detectors at the relay station. Ref.~\cite{RELAY} found
that the optimal performances are achieved in the asymmetric configuration
where the relay is close to one of the parties.

In this paper, we deepen the analysis of Ref.~\cite{RELAY} considering the
symmetric configuration where the relay is midway between Alice and Bob.
Despite the fact this is not the optimal setup in terms of rates and security
distances, it is still very important to analyze for potential applications in
network scenarios where two parties are roughly equidistant from a public
server or access point. Another reason for analyzing this kind of setup is in
its simple analytical formulas for the secret-key rates. Using these formulas,
we can provide a very detailed comparison between the most important Gaussian
attacks against the quantum links.

In our cryptanalysis of the symmetric protocol we clearly show that, at fixed
transmissivity and thermal noise on the links, the optimal eavesdropping
strategy is a two-mode coherent Gaussian attack, where Eve injects quantum
correlations in both the quantum channels leading to the relay. In this
attack, these channels are combined (via beam-splitters) with two entangled
modes, which have been prepared in a suitable Einstein-Podolsky-Rosen (EPR)
state. Such a strategy greatly outperforms the single-mode collective attack
based on two independent entangling cloners which was assumed in some recent
investigations \cite{Li,Ma,NoteEXCESS}. Any such security analysis relying on
independent attacks on the channels is therefore incomplete and opens security loopholes.

The paper is organized as follows: In Section~\ref{PROT} we describe the setup
in the symmetric scenario. In Section~\ref{EAVESDROPPING} we analyze its
security and provide a formula for the key rate. In Section~\ref{Sym_APP} we
compare the various Gaussian attacks, identifying the optimal attack and the
corresponding minimum key rate of the protocol. Finally, in Section~\ref{CONC}
we draw our conclusions.

\section{The protocol\label{PROT}}

\begin{figure}[t]
\vspace{+0.0cm}
\par
\begin{center}
\includegraphics[width=0.45\textwidth] {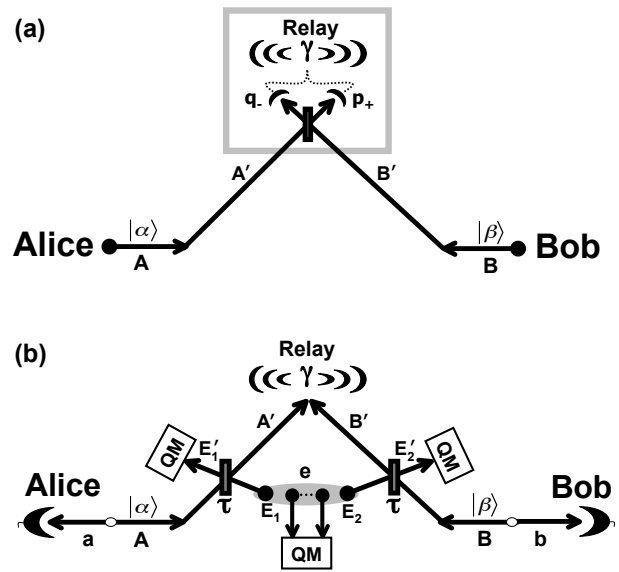}
\end{center}
\par
\vspace{-0.6cm}\caption{(Color online) (a) Relay-based protocol performed in
the symmetric configuration with the untrusted relay perfectly in the middle
between the parties. Alice and Bob prepare coherent states with
Gaussianly-modulated amplitudes, $\alpha$ and $\beta$, respectively. The relay
performs a continuous-variable Bell measurement with complex outcome
$\gamma:=(q_{-}+ip_{+})/\sqrt{2}$, which is publicly broadcast. From the
knowledge of $\gamma$ each party can infer the variable of the other party.
(b) Entanglement-based representation of the protocol under two-mode Gaussian
attack. Using two beam splitters with transmissivity $\tau$, Eve injects to
ancillary modes, $E_{1}$ and $E_{2}$, prepared in a two-mode Gaussian state
with zero mean and CM in the symmetric normal form of Eq.~(\ref{EVE-CM}). See
text for details.}%
\label{scheme}%
\end{figure}

Let us consider the scenario where Alice and Bob do not access a direct
communication link. Instead they connect to a perfectly-in-the-middle relay
via insecure quantum links, as shown in Fig. \ref{scheme}(a). The relay is
untrusted, meaning that it is assumed to be operated by Eve in the worst case scenario.

The protocol proceeds as follows: Alice and Bob possess two modes, $A$ and $B
$, which are prepared in coherent states, $|\alpha\rangle$ and $|\beta\rangle
$, with randomly-modulated amplitudes (according to a complex Gaussian
distribution with large variance). They send these modes to the intermediate
relay where the output modes, $A^{\prime}$ and $B^{\prime}$, are subject to a
continuous-variable Bell detection~\cite{BellFORMULA}. This means that
$A^{\prime}$ and $B^{\prime}$ are mixed on a balanced beam splitter and the
output ports are conjugately homodyned: One port is homodyned in the $\hat{q}
$-quadrature with outcome $q_{-}$, while the other port is homodyned in the
$\hat{p}$-quadrature with outcome $p_{+}$. Compactly, the measurement provides
the complex outcome $\gamma:=(q_{-}+ip_{+})/\sqrt{2}$ which is then broadcast
over a public channel. 

To understand the working mechanism of the relay, first suppose there is no
loss and noise in the links. In such a case, we have $\gamma\simeq\alpha
-\beta^{\ast}$, so that the communication of $\gamma$ creates \textit{a
posteriori} correlations between Alice's and Bob's variables. Each party can
easily infer the variable of the other. For instance, Alice could compute the
quantity $\alpha-\gamma\simeq\beta^{\ast}$ recovering Bob's encoding $\beta$
up to detection noise~\cite{RELAY}. This procedure partly recalls the
post-processing of the two-way QKD\ protocols~\cite{TwoWay,Weedbrook2013}.

Note that Eve's knowledge of variable $\gamma$ would be of no help to extract
information on the individual variables $\alpha$ and $\beta$, i.e., we have
$I(\alpha:\gamma)=I(\beta:\gamma)=0$ in terms of mutual information
\cite{Nielsen,Wilde}. By contrast, as a result of the broadcast of $\gamma$,
the conditional mutual information of Alice and Bob becomes non-zero
$I(\alpha:\beta|\gamma)>I(\alpha:\beta)$. Thus, if Eve wants to steal
information, she needs to introduce loss and noise.

In general, the action of Eve may involve a global unitary operation
correlating all the uses of the protocols. However, using random permutations
of their data~\cite{Renner,renner-cirac}, Alice and Bob can always reduce this
scenario to an attack which is coherent within the single use of protocol.
This can be a joint attack of both the links and the relay. The parties can
further reduce this eavesdropping to consider a coherent attack of the links
only, assuming a properly-working relay, i.e., a relay implementing a
continuous-variable Bell detection. In particular, since the protocol is based
on the Gaussian modulation and Gaussian detection of Gaussian states, the
optimal coherent attack of the links will be based on a Gaussian unitary
interaction~\cite{RMP,patron-gaussian,navascues}. See Ref.~\cite{RELAY} for
more details about this general reduction of the attack.

\section{Cryptoanalysis of the protocol\label{EAVESDROPPING}}

According to the previous discussion, we can reduce the attack to a two-mode
Gaussian attack against the quantum links of Alice and\ Bob. The most
realistic implementation of such an attack consists of two beam splitters
combining the signals with ancillary modes prepared in a generally
quantum-correlated Gaussian state (more exotic Gaussian attacks may be
constructed using other canonical forms~\cite{canATTACKS}). In particular, we
will consider the symmetric configuration of this attack, where the parameters
are identical for the two links, so that the performances of the protocol are
invariant under exchange of Alice and Bob. In order to analyze this symmetric
Gaussian attack, we adopt the entanglement-based (EB) representation of the
protocol, where Alice's and Bob's ensembles of coherent states are simulated
using two EPR states subject to local heterodyne detections.

\subsection{Entangled-based representation}

The EB-representation of the protocol is described in Fig.~\ref{scheme}(b).
Alice and Bob possess two EPR states, $\rho_{aA}$ and $\rho_{bB}$, with the
same covariance matrix (CM)~\cite{RMP}%
\begin{equation}
\mathbf{V}(\mu)=\left(
\begin{array}
[c]{cc}%
\mu\mathbf{I} & \sqrt{\mu^{2}-1}\mathbf{Z}\\
\sqrt{\mu^{2}-1}\mathbf{Z} & \mu\mathbf{I}%
\end{array}
\right)  ,
\end{equation}
where $\mathbf{I}=\mathrm{diag}(1,1)$, $\mathbf{Z}=\mathrm{diag}(1,-1)$ and
$\mu\geq1$.

By applying heterodyne detections to the remote modes, $a$ and $b$, Alice and
Bob prepares coherent states, $|\alpha\rangle$ and $|\beta\rangle$, on the
other EPR\ modes, $A$ and $B$, respectively. In particular, their amplitudes,
$\alpha$ and $\beta$, are modulated according to a complex Gaussian
distribution with variance $\varphi:=\mu-1$, that we take to be very large
$\varphi\gg1$. Modes $A$ and $B$ are then sent to the relay for detection.

\subsection{Symmetric two-mode Gaussian eavesdropping}

Alice's and Bob's modes $A$ and $B$ are mixed with ancillary modes, $E_{1}$
and $E_{2}$, respectively.\ This is done by means of two beam-splitters with
the same transmissivity $\tau$. The ancillas belong to an environmental set
$\{E_{1},E_{2},\mathbf{e}\}$ in the hands of Eve, and the reduced state of
$E_{1}$ and $E_{2}$ is a zero-mean Gaussian state $\sigma_{E_{1}E_{2}}$ with
CM in the symmetric normal form
\begin{equation}
\mathbf{V}_{E_{1}E_{2}}=\left(
\begin{array}
[c]{cc}%
\omega\mathbf{I} & \mathbf{G}\\
\mathbf{G} & \omega\mathbf{I}%
\end{array}
\right)  ,~\mathbf{G}:=\left(
\begin{array}
[c]{cc}%
g & 0\\
0 & g^{\prime}%
\end{array}
\right)  . \label{EVE-CM}%
\end{equation}
Here $\omega$ is the variance of the thermal noise injected in the beam
splitters, while $\mathbf{G}$\ accounts for the quantum correlations between
the two ancillas. (The various parameters $\omega$, $g$, and $g^{\prime}$
satisfy simple physical constraints imposed by the uncertainty
principle~\cite{TwomodePRA,NJP2013}).

The output modes, $A^{\prime}$ and $B^{\prime}$, are subject to the
continuous-variable Bell detection (with the outcome broadcast), while Eve's
output modes, $E_{1}^{\prime}$ and $E_{2}^{\prime}$, together with all the
other ancillary modes $\mathbf{e}$ are stored in a quantum memory, which is
detected by an optimal coherent measurement at the end of the protocol.

Note that we may consider different transmissivities, $\tau_{A}$ and $\tau
_{B}$, for the beam splitters, and an asymmetric CM with different thermal
variances, $\omega_{A}$ and $\omega_{B}$. This is the general asymmetric case
considered in Ref.~\cite{RELAY}. However, when the relay is midway between the
two parties, the amount of loss and noise present in the links is
realistically expected to be identical. In other words, it is reasonable to
consider here a symmetric attack as the one previously described, which has
\begin{equation}
\tau_{A}=\tau_{B}:=\tau,~\omega_{A}=\omega_{B}:=\omega~. \label{Sym}%
\end{equation}

Thanks to this symmetry, we can reduce the number of parameters and derive a
simple analytical expression for the secret-key rate, which allows us to
perform a detailed analysis of the various specific symmetric attacks which
are possible against our protocol. In particular, we can easily study the
performances of these attacks in terms of the correlation parameters, $g$ and
$g^{\prime}$, and identify the optimal one which minimizes key rate and
security threshold. Furthermore, due to the symmetry, Alice and Bob can be
interchanged, which implies that there is no difference between direct and
reverse reconciliation~\cite{RMP}. In other words, we can consider a unique
secret-key rate for the protocol (assuming one-way classical communication for
error correction and privacy amplification).

\subsection{Secret-key rate}

Without loss of generality, we assume that Alice is the encoder of information
(variable $\alpha$) while Bob is the decoder, so that he post-processes his
variable $\beta$ to infer $\alpha$. In the EB-representation, these variables
are informationally equivalent to the outcomes of the heterodyne
detections~\cite{RELAY}. To derive the rate, we note that the Bell detection
at the relay and the heterodyne detections of the two parties commute with
each other. For this reason, we can equivalently compute the rate from the
conditional state $\rho_{ab|\gamma}$ of modes $a$ and $b$ after the
communication of the outcome $\gamma$. The rate is given by~\cite{RELAY}%
\begin{equation}
R=I_{ab|\gamma}-I_{E|\gamma}, \label{RATE-GEN}%
\end{equation}
where $I_{ab|\gamma}$ is Alice and Bob's conditional mutual information, while
$I_{E|\gamma}$ is Eve's Holevo information~\cite{Wilde} on Alice's variable
(which can be computed from the state of the output ancillas).

Since all output modes are in a global pure state and the various detections
are rank-1, we can write~\cite{RELAY}%
\begin{equation}
I_{E|\gamma}=S(\rho_{ab|\gamma})-S(\rho_{b|\gamma\alpha}), \label{IEgamma}%
\end{equation}
where $S(.)$ is the von Neumann entropy~\cite{Wilde}, computed on the
post-relay state $\rho_{ab|\gamma}$ of Alice and Bob, and the
double-conditional state $\rho_{b|\gamma\alpha}$ of Bob, conditioned to
relay's and Alice's detections (computable from $\rho_{ab|\gamma}$).

\subsection{Computation of the key rate\label{sec:RATE}}

Both the mutual information $I_{ab|\gamma}$ and Eve's Holevo entropy
$I_{E|\gamma}$ can be computed from the post-relay state $\rho_{ab|\gamma}$,
in particular, from its CM $\mathbf{V}_{ab|\gamma}$. Imposing the symmetry
conditions of Eq.~(\ref{Sym}) in the general expression of $\mathbf{V}%
_{ab|\gamma}$ computed in Ref.~\cite{RELAY}, we derive%
\begin{align}
\mathbf{V}_{ab|\gamma}  &  =\left(
\begin{array}
[c]{cc}%
\mu\mathbf{I} & \mathbf{0}\\
\mathbf{0} & \mu\mathbf{I}%
\end{array}
\right)  -\frac{\tau(\mu^{2}-1)}{2}\times\label{totalCM-PR}\\
&  \times\left(
\begin{array}
[c]{cccc}%
\frac{1}{\tau\mu+\lambda} &  & -\frac{1}{\tau\mu+\lambda} & \\
& \frac{1}{\tau\mu+\lambda^{\prime}} &  & \frac{1}{\tau\mu+\lambda^{\prime}}\\
-\frac{1}{\tau\mu+\lambda} &  & \frac{1}{\tau\mu+\lambda} & \\
& \frac{1}{\tau\mu+\lambda^{\prime}} &  & \frac{1}{\tau\mu+\lambda^{\prime}}%
\end{array}
\right)  ,\nonumber
\end{align}
where
\begin{equation}
\lambda:=(1-\tau)(\omega-g),~\lambda^{\prime}:=(1-\tau)(\omega+g^{\prime}).
\label{lambdas}%
\end{equation}
Note that Eq. (\ref{totalCM-PR}) represents a particular case of the general
CM of Eq. (\ref{VabCondGammaGEN2}), which is obtained in the Appendix, where
non-unit quantum efficiencies of the detectors are also included.

Now, we can easily compute \cite{RMP} the symplectic spectrum of
eq.(\ref{totalCM-PR}) in the limit of large modulation $\mu\gg1$, obtaining%
\begin{equation}
\nu_{1}\rightarrow\sqrt{\frac{\lambda\mu}{\tau}}~,\nu_{2}\rightarrow
\sqrt{\frac{\lambda^{\prime}\mu}{\tau}}. \label{total-SPECTRUM}%
\end{equation}
Then, entropy term $S(\rho_{ab|\gamma})$ in Eq.~(\ref{IEgamma}) can be
computed using the function%
\begin{align}
h(x)  &  :=\frac{x+1}{2}\log_{2}\frac{x+1}{2}-\frac{x-1}{2}\log_{2}\frac
{x-1}{2}\label{H}\\
&  \rightarrow\log\frac{e}{2}x+O\left(  \frac{1}{x}\right)  ~~\text{for }%
x\gg1. \label{H-asy}%
\end{align}
Thus, we have
\begin{equation}
S(\rho_{ab|\gamma})=h(\nu_{1})+h(\nu_{2})\rightarrow\log\frac{e^{2}}{4\tau
}\sqrt{\lambda\lambda^{\prime}}\mu. \label{totENT}%
\end{equation}

To compute $S(\rho_{b|\gamma\alpha})$ we derive the double-conditional CM
$\mathbf{V}_{b|\gamma\alpha}$. We put Eq.~(\ref{totalCM-PR}) in the block-form%
\begin{equation}
\mathbf{V}_{ab|\gamma}=\left(
\begin{array}
[c]{cc}%
\mathbf{A} & \mathbf{C}\\
\mathbf{C}^{T} & \mathbf{B}%
\end{array}
\right)  ,
\end{equation}
and we apply a partial gaussian heterodyne measurement on Alice's remote mode
$a,$given by \cite{RMP,giedke,BellFORMULA},
\begin{equation}
\mathbf{V}_{b|\gamma\alpha}=\mathbf{B-C}^{T}(\mathbf{A+I)}^{-1}\mathbf{C},
\label{HET-FORMULA}%
\end{equation}
which gives%
\begin{equation}
\mathbf{V}_{b|\gamma\alpha}=\left(
\begin{array}
[c]{cc}%
\mu-\frac{(\mu^{2}-1)\tau}{\tau(\mu+1)+2\lambda} & 0\\
0 & \mu-\frac{(\mu^{2}-1)\tau}{\tau(\mu+1)+2\lambda^{\prime}}%
\end{array}
\right)  .\label{bob-CM-cond}%
\end{equation}
For $\mu\gg1$, its symplectic eigenvalue is given by%
\begin{equation}
\nu\rightarrow\frac{\sqrt{(\tau+2\lambda)(\tau+2\lambda^{\prime})}}{\tau},
\label{COND-SPECTRUM}%
\end{equation}
and we have $S(\rho_{b|\gamma\alpha})=h(\nu)$. We can then compute Eve's
Holevo information, asymptotically given by%
\begin{equation}
I_{E|\gamma}=\log_{2}\frac{e^{2}\sqrt{\lambda\lambda^{\prime}}\mu}{4\tau
}-h\left[  \frac{\sqrt{(\tau+2\lambda)(\tau+2\lambda^{\prime})}}{\tau}\right]
. \label{HOLEVO-BOUND}%
\end{equation}
Alice and Bob's conditional mutual information $I_{ab|\gamma}$ can be computed
from the classical CM\ $\mathbf{V}(\alpha,\beta|\gamma)=(\mathbf{V}%
_{ab|\gamma}+\mathbf{I})/2$ describing their outcomes. After simple algebra,
we get the asymptotic expression%
\begin{equation}
I_{AB|\gamma}=\log_{2}\frac{\tau\mu}{4\sqrt{(\tau+\lambda)(\tau+\lambda
^{\prime})}}. \label{IAB}%
\end{equation}
As a result, we computed the following asymptotic secret-key rate for the
symmetric Gaussian attack%
\begin{align}
R_{sym}  &  =\log_{2}\left[  \frac{\tau^{2}}{e^{2}\sqrt{\lambda\lambda
^{\prime}(\tau+\lambda)(\tau+\lambda^{\prime})}}\right] \nonumber\\
&  +h\left[  \frac{\sqrt{(\tau+2\lambda)(\tau+2\lambda^{\prime})}}{\tau
}\right]  , \label{Rsym}%
\end{align}
which is function of the parameters $\tau$, $\omega$, $g$ and $g^{\prime}$.

A complete analysis of the performances of the scheme in presence of non-ideal
experimental conditions is described in Appendix \ref{app-efficiency}

\section{Detailed analysis of the symmetric attacks\label{Sym_APP}}

\begin{figure}[t]
\vspace{+0.2cm}
\par
\begin{center}
\includegraphics[width=0.32\textwidth] {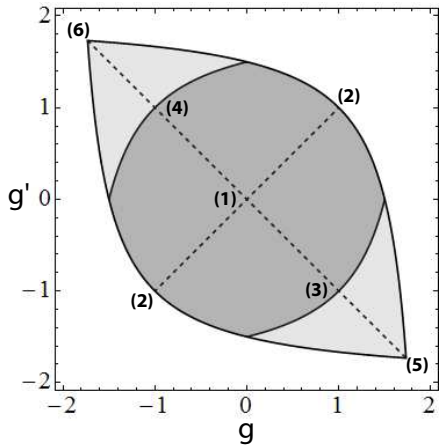}
\end{center}
\par
\vspace{-0.6cm}\caption{(Color online) Correlation plan for a symmetric
Gaussian attack. Given $\tau$ and $\omega$ (here set to $2$), the attack is
fully specified by the two correlation parameters $(g,g^{\prime})$, whose
accessible values are represented by the non-white area. In particular, the
inner darker region represents the set of separable attacks ($\sigma
_{E_{1}E_{2}}$ separable), while the two outer and lighter regions represent
entangled attacks ($\sigma_{E_{1}E_{2}}$ entangled). The numbered points
represent the specific attacks described in Sec.~\ref{Sym_APP}.}%
\label{AttackCHA}%
\end{figure}

For fixed transmissivity $\tau$ and thermal noise $\omega$ affecting each
link, there are remaining degrees of freedom in the two-mode Gaussian attack.
These are given by the correlation parameters $g$ and $g^{\prime}$, which can
be represented as a point on a `correlation plane' (see Fig.~\ref{AttackCHA}).
Each point of this plane describes an attack (with different amount and kind
of correlations) to which it corresponds a specific key rate according to
Eq.~(\ref{Rsym}). Here we provide a detailed comparison between these attacks,
showing how the optimal coherent attack greatly outperforms the collective
attack based on independent entangling cloners.

Because of the symmetry, we have a simple characterization of the set of
possible Gaussian attacks which are accessible to Eve. These correspond to
points $(g,g^{\prime})$ such that~\cite{NJP2013,TwomodePRA}%
\begin{align}
|g|  &  <\omega,~|g^{\prime}|<\omega,\\
\omega\left\vert g+g^{\prime}\right\vert  &  \leq\omega^{2}+gg^{\prime}-1.
\end{align}
Among all these accessible attacks, those satisfying the further condition%
\begin{equation}
\omega^{2}-gg^{\prime}-1\geq\omega|g-g^{\prime}| \label{sepCON}%
\end{equation}
are separable attacks ($\sigma_{E_{1}E_{2}}$ separable), while those violating
Eq.~(\ref{sepCON}) are entangled attacks ($\sigma_{E_{1}E_{2}}$ entangled).
See Fig.~\ref{AttackCHA} for a numerical representation. In particular, we can identify the following attacks:

\emph{Collective attack. }This is the simplest attack, represented by
point~(1) in Fig.~\ref{AttackCHA}, i.e., the origin of the plane
($g=g^{\prime}=0$). This corresponds to using two identical and independent
entangling cloners with transmissivity $\tau$ and thermal noise $\omega$. In
fact, we have $\sigma_{E_{1}E_{2}}=\sigma_{E_{1}}\otimes\sigma_{E_{2}}$, where
$\sigma_{E_{k}}$ ($k=1,2$) is a thermal state with variance $\omega$, whose
purification $\Phi_{E_{k}e_{k}}$ is an EPR state in the hands of Eve.

\emph{Separable attacks\textbf{.}} Within the separable attacks we can
identify points~(2), (3), and~(4) in Fig.~\ref{AttackCHA}. These are
characterized by the condition $|g|=|g^{\prime}|=\omega-1$ and represent the
separable attacks with the highest correlations. In particular, points~(2)
correspond to the cases $g=g^{\prime}=\omega-1$ or $g=g^{\prime}=1-\omega$,
point~(3) corresponds to $g=-g^{\prime}=\omega-1$, and point~(4) to
$g=-g^{\prime}=1-\omega$.

\emph{EPR attacks.} Finally, points (5) and (6) in Fig.~\ref{AttackCHA} are
the most entangled attacks, where Eve's ancillas $E_{1}$ and $E_{2}$ are
described by an EPR state. Point (5) is the `positive EPR attack' with
$g=-g^{\prime}=\sqrt{\omega^{2}-1}$, while (6) is the `negative EPR attack'
with $g=-g^{\prime}=-\sqrt{\omega^{2}-1}$. The latter turns out to be the
optimal attack against the protocol.

\begin{figure}[t]
\vspace{+0.2cm}
\par
\begin{center}
\includegraphics[width=0.45\textwidth] {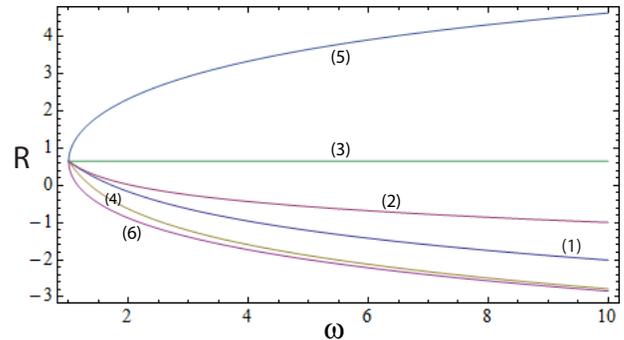}
\end{center}
\par
\vspace{-0.7cm}\caption{(Color online) Secret-key rate $R$ (bits) versus
thermal noise $\omega$\ for the various symmetric attacks (1)-(6) classified
in Sec.~\ref{Sym_APP} and displayed in Fig.~\ref{AttackCHA}.
Link-transmissivity is set to $\tau=0.9$. Note that the negative EPR\ attack
(6) is the optimal attack minimizing the rate of the protocol. }%
\label{RatesCOMP}%
\end{figure}

\begin{figure}[t]
\par
\begin{center}
\includegraphics[width=0.47\textwidth] {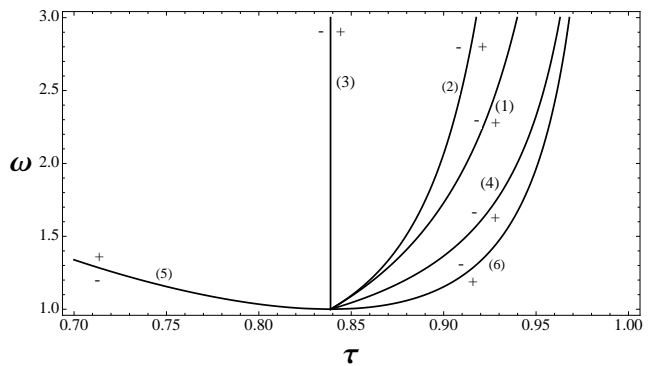}
\end{center}
\par
\vspace{-0.7cm}\caption{(Color online) Security threshold ($R=0$) expressed as
maximum tolerable thermal noise $\omega$\ versus link-transmissivity $\tau$.
We compare the various symmetric attacks (1)-(6) classified in
Sec.~\ref{Sym_APP} and displayed in Fig.~\ref{AttackCHA}. The negative EPR
attack (6) is the optimal corresponding to the lowest security threshold. Also
note the peculiar inversion of the threshold for the positive EPR attack (5),
for which the rate is positive for values of thermal noise \textit{above} the
threshold.}%
\label{soglie}%
\end{figure}

We now compare the performances of the previous attacks in
Figs.~\ref{RatesCOMP} and~\ref{soglie}. In Fig.~\ref{RatesCOMP} we fix the
transmissivity $\tau=0.9$ and we study the corresponding rates $R$ as function
of the thermal noise $\omega$. In Fig.~\ref{soglie} we plot the security
thresholds. These are given by the condition $R=0$, and they are expressed in
terms of maximum tolerable thermal noise versus transmissivity $\omega
=\omega(\tau)$. 

Our analysis identifies \textquotedblleft good\textquotedblright\ and
\textquotedblleft bad\textquotedblright\ entanglement for the security of the
protocol. Good entanglement refers to the entangled attacks in the bottom
right area of Fig.~\ref{AttackCHA}, with $g=-g^{\prime}>0$, of which the
attacks (3) and (5) are border points. This entanglement is good because it
injects correlations of the type $\hat{q}_{E_{1}}\approx\hat{q}_{E_{2}}$ and
$\hat{p}_{E_{1}}\approx-\hat{p}_{E_{2}}$, therefore helping the Bell detection
(which projects on $\hat{q}_{A^{\prime}}\approx\hat{q}_{B^{\prime}} $ and
$\hat{p}_{A^{\prime}}\approx-\hat{p}_{B^{\prime}}$). As a result, Eve actively
helps the key distribution.

This is evident from the performance of the positive EPR attack (5) both in
terms of rate (Fig.~\ref{RatesCOMP}) and security threshold (Fig.~\ref{soglie}%
). In fact, from Fig.~\ref{RatesCOMP}, we see that the rate is
\textit{increasing} in the thermal noise $\omega$ and, in Fig.~\ref{soglie},
we see a peculiar inversion of the security threshold so that thermal noise
\textit{above} the threshold is tolerable. These features are typical of all
entangled attacks with $\omega-1<g\leq\sqrt{\omega^{2}-1}$ and $g^{\prime}=-g
$, corresponding to the segment of points between (3) (excluded) and (5)
(included). In Figs.~\ref{RatesCOMP} and~\ref{soglie}, these attacks have
curves which are intermediate between those of (3) and (5).

By contrast, bad entanglement refers to the entangled attacks in the top left
area of Fig.~\ref{AttackCHA}, with $g=-g^{\prime}<0$ and having the attacks
(4) and (6) as border points. This entanglement is instead bad because it
injects correlations of the type $\hat{q}_{E_{1}}\approx-\hat{q}_{E_{2}}$ and
$\hat{p}_{E_{1}}\approx\hat{p}_{E_{2}}$, which are opposite to those
established by the Bell detection. In this case, Eve decreases the
correlations between Alice's and Bob's variables, and she is able to eavesdrop
more information, with the optimal strategy achieved by the negative EPR
attack (6) as clear from the rates of Fig.~\ref{RatesCOMP} and the security
thresholds of Fig.~\ref{soglie}. The asymmetric version of this attack is
optimal in case of asymmetric setups~\cite{RELAY}.

By comparing the curves (6) and (1) in Figs.~\ref{RatesCOMP} and~\ref{soglie},
we clearly see the substantial advantage given by this optimal attack with
respect to the standard collective attack based on independent entangling
cloners. Analytically, the minimum key rate associated with the optimal attack
is given by%
\begin{equation}
R_{\min}=h\left(  \frac{\tau+2\lambda_{\text{opt}}}{\tau}\right)  +\log
_{2}\left[  \frac{\tau^{2}}{e^{2}\lambda_{\text{opt}}(\tau+\lambda
_{\text{opt}})}\right]  , \label{Rmin}%
\end{equation}
with $\lambda_{\text{opt}}=(1-\tau)(\omega+\sqrt{\omega^{2}-1})$. One can
easily check this is numerically much less than the rate of the collective
attack%
\begin{align}
R_{\text{coll}}  &  =h\left[  \frac{\tau+2(1-\tau)\omega}{\tau}\right]
\nonumber\\
&  +\log_{2}\left\{  \frac{\tau^{2}}{e^{2}(1-\tau)[\tau+(1-\tau)\omega]\omega
}\right\}  .
\end{align}

Thus, the security analysis which is valid for one-way continuous-variable QKD
protocols~\cite{RMP}, and based on the study of collective (single-mode)
entangling-cloner attacks, cannot be applied to our relay-based protocol. For
this reason, the studies provided by Refs. \cite{Li,Ma} are incomplete and
cannot prove the unconditional security of the relay-based (measurement-device
independent) QKD with continuous variables.

\section{Conclusions\label{CONC}}

In conclusion, we have provided a detailed analysis of the relay-based QKD
protocol of Ref.~\cite{RELAY}, considering a completely symmetric setup under
the action of symmetric attacks. Despite the fact that this particular case
does not represent the optimal configuration of the scheme, still it is
important for its potential implementation in network scenarios. Furthermore,
the symmetry conditions allow us to greatly simplify the cryptoanalysis and
derive simple analytical results.

Thanks to this approach, we have been able to provide a very detailed
classification of the symmetric attacks against the protocol, characterizing
the different possible strategies in terms of their correlation properties. At
fixed transmissivity and thermal noise, we have identified the optimal
symmetric attack which corresponds to a coherent attack where the ancillas are
maximally entangled with a specific type of EPR\ correlations (negative EPR
attack). In particular, this attack greatly outperforms the standard
collective attack based on independent entangling-cloners, which is therefore
unsuitable for assessing the security of this kind of protocol (contrary to
the claims of other analysis \cite{Li, Ma}). This also confirms the results of
Ref.~\cite{RELAY} which were given in the general case of asymmetric setups.

Finally our work clarifies the crucial role of quantum correlations in
assessing the security of QKD\ protocols which are based on untrusted relays.
Further studies may include the extension of this protocol to
thermal-QKD~\cite{Weedbrook2010,Weedbrook2012,Weedbrook2013}, with the aim of
using different wavelengths of the electromagnetic field, e.g., in mixed
technology platforms using both optical/infrared and microwave carriers.

\section*{Acknowledgments}

This work was funded by a Leverhulme Trust research fellowship, the EPSRC via
`HIPERCOM' (grant no. EP/J00796X/1), `qDATA' (grant no. EP/L011298/1) and the
UK Quantum Technology Hub for Quantum Communications Technologies (Grant no. EP/M013472/1).

\appendix

\section{Extension to experimental imperfections}

\label{app-efficiency}In this appendix we analyze the role of experimental
imperfections computing the key-rates and the security thresholds in the
presence of Bell detectors with non-ideal quantum efficiencies. We also study
finite-size effects connected with finite Gaussian modulations~\cite{RMP}, and
the role played by the non-ideal efficiency of the classical reconciliation
codes~\cite{leverrier}. We show that, also in the presence of realistic
experimental limitations, the optimal eavesdropping is given by the two-mode
coherent \textquotedblleft negative-EPR attack\textquotedblright.

\subsection{Post-relay covariance matrix for non-ideal Bell detectors}

\begin{figure}[t]
\vspace{+0.2cm}
\par
\begin{center}
\includegraphics[width=0.25\textwidth] {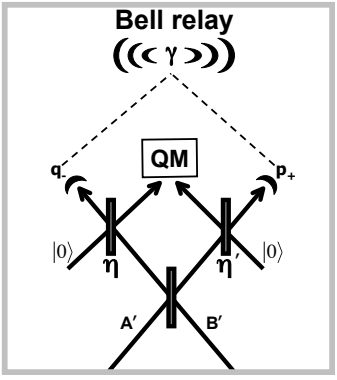}
\end{center}
\par
\vspace{-0.4cm}\caption{(Color online) Untrusted relay with inefficient
detectors. Two additional beam splitters, with transmissivities $(\eta
,\eta^{\prime})$ are placed in front of the ideal detectors. One output from
each beam splitter is sent to the detectors for measurements. The other
outputs are sent to Eve's quantum memory.}%
\label{relay_eff}%
\end{figure}

We generalize Eq.(\ref{totalCM-PR}) to include detectors' efficiencies by
placing two beam splitters with transmissivities $\eta$ and $\eta^{\prime}$,
as illustrated in Fig. \ref{relay_eff}. To preserve the purity of the global
(Alice-Bob-Eve) state, the non-detected signals are sent to Eve's quantum
memory (this is the assumption to make in the worst-case scenario, since the
relay is untrusted and Eve can control the loss of the detectors).

We follow the general approach given in Ref. \cite{BellFORMULA}. We write the
total CM in the block form%
\begin{equation}
\mathbf{V}=\left(
\begin{array}
[c]{cc}%
\mathbf{V}_{ab} & \mathbf{C}\\
\mathbf{C}^{T} & \mathbf{B}%
\end{array}
\right)  , \label{Block-general-CM}%
\end{equation}
where the block
\begin{equation}
\mathbf{B=}\left(
\begin{array}
[c]{cc}%
\mathbf{B}_{1} & \mathbf{D}\\
\mathbf{D}^{T} & \mathbf{B}_{2}%
\end{array}
\right)  , \label{B-appendix}%
\end{equation}
describes the modes sent to the relay, $A^{\prime}$ and $B^{\prime}$. These
are processed by the balanced beam splitter and then measured. In our case it
is easy to verify~\cite{RELAY,BellFORMULA} that the blocks $\mathbf{B}_{1}
$,$\mathbf{B}_{2}$ and $\mathbf{D}$ take the following expressions%
\begin{align}
\mathbf{B}_{1}  &  =\mathbf{B}_{2}=[\tau\mu+(1-\tau)\omega]\mathbf{I,}%
\label{B1-B2}\\
\mathbf{D}  &  \mathbf{=}(1-\tau)\mathbf{G,} \label{D-appendix}%
\end{align}
where $\mathbf{I}=$\textrm{diag}$(1,1)$ and $\mathbf{G}=$\textrm{diag}%
$(g,g^{\prime}).$

In Eq. (\ref{Block-general-CM}) the sub-matrix $\mathbf{V}_{ab}$ describes the
joint quantum state of remote modes $a$ and $b$, while the block
$\mathbf{C=(C}_{1}\mathbf{C}_{2}\mathbf{)}$\textbf{\ }is a rectangular matrix
accounting for the correlations between the remote modes and the transmitted
ones, i.e., $A^{\prime}$ and $B^{\prime}$. In particular, we compute%
\begin{equation}
\mathbf{C}_{1}=\left(
\begin{array}
[c]{c}%
\sqrt{\tau(\mu^{2}-1)}\mathbf{Z}\\
\mathbf{0}%
\end{array}
\right)  ,\text{ }\mathbf{C}_{2}=\left(
\begin{array}
[c]{c}%
\mathbf{0}\\
\sqrt{\tau(\mu^{2}-1)}\mathbf{Z}%
\end{array}
\right)  . \label{C1-C2}%
\end{equation}

Applying Eq.~(74) from Ref.~\cite{BellFORMULA} to Eq.~(\ref{Block-general-CM}%
), we obtain Alice and Bob's CM conditioned to the relay Bell measurement.
This is given by%
\begin{equation}
\mathbf{V}_{ab|\gamma}=\mathbf{V}_{ab}-\frac{1}{2\det\mathbf{\gamma(}\eta
,\eta^{\prime}\mathbf{)}}%
{\displaystyle\sum\limits_{i,j=1,2}}
\mathbf{C}_{i}\left(  \mathbf{X}_{i}^{T}\mathbf{\gamma\mathbf{(}}\eta
,\eta^{\prime}\mathbf{\mathbf{)}X}_{j}\right)  \mathbf{C}_{j}^{T},
\label{VabCondGammaGEN}%
\end{equation}
where%
\begin{equation}
\mathbf{X}_{1}=\left(
\begin{array}
[c]{cc}
& 1\\
1 &
\end{array}
\right)  ,\text{ }\mathbf{X}_{2}=\left(
\begin{array}
[c]{cc}
& 1\\
-1 &
\end{array}
\right)  . \label{X1-X2}%
\end{equation}
Here the quantum efficiencies of the detectors, simulated by the beam splitter
transmissivities $\eta$ and $\eta^{\prime}$, are contained in the symmetric
matrix%
\begin{equation}
\mathbf{\gamma(}\eta,\eta^{\prime}\mathbf{)=}\left(
\begin{array}
[c]{cc}%
\gamma_{1}(\eta) & \gamma_{3}\\
\gamma_{3} & \gamma_{2}(\eta^{\prime})
\end{array}
\right)  . \label{gamma-M}%
\end{equation}

In the case of a Bell detection the matrix $\mathbf{\gamma(}\eta,\eta^{\prime
}\mathbf{)}$ can explicitly be computed, according to Eqs.~(54-59) of
Ref.~\cite{BellFORMULA}. In particular, its entries take the form%
\begin{align}
\gamma_{1}(\eta)  &  =\gamma_{1}+\frac{1-\eta}{\eta},\nonumber\\
\gamma_{2}(\eta^{\prime})  &  =\gamma_{2}+\frac{1-\eta^{\prime}}{\eta^{\prime
}},\nonumber\\
\gamma_{3}  &  =0 \label{gammas}%
\end{align}
where $\gamma_{1}=\tau\mu+(1-\tau)(\omega-g)$ and $\gamma_{2}=\tau\mu
+(1-\tau)(\omega+g^{\prime})$ are easily obtained from Eq.~(\ref{B1-B2})
and~(\ref{D-appendix}), using the formulas of Ref.~\cite{BellFORMULA}.


After simple algebra we derive the post-relay CM inclusive of the quantum
efficiencies%
\begin{align*}
\mathbf{V}_{ab|\gamma}  &  =\left(
\begin{array}
[c]{cc}%
\mu\mathbf{I} & \mathbf{0}\\
\mathbf{0} & \mu\mathbf{I}%
\end{array}
\right)  -\frac{\tau(\mu^{2}-1)}{2}\times\\
&  \times\left(
\begin{array}
[c]{cccc}%
\frac{1}{\gamma_{1}(\eta)} &  & -\frac{1}{\gamma_{1}(\eta)} & \\
& \frac{1}{\gamma_{2}(\eta^{\prime})} &  & \frac{1}{\gamma_{2}(\eta^{\prime}%
)}\\
-\frac{1}{\gamma_{1}(\eta)} &  & \frac{1}{\gamma_{1}(\eta)} & \\
& \frac{1}{\gamma_{2}(\eta^{\prime})} &  & \frac{1}{\gamma_{2}(\eta^{\prime})}%
\end{array}
\right)  .
\end{align*}
This CM can be rewritten in the form%
\begin{align}
\mathbf{V}_{ab|\gamma}  &  =\left(
\begin{array}
[c]{cc}%
\mu\mathbf{I} & \mathbf{0}\\
\mathbf{0} & \mu\mathbf{I}%
\end{array}
\right)  -\frac{\tau(\mu^{2}-1)}{2}\times\nonumber\\
&  \times\left(
\begin{array}
[c]{cccc}%
\frac{1}{\tau\mu+\lambda(\eta)} &  & -\frac{1}{\tau\mu+\lambda(\eta)} & \\
& \frac{1}{\tau\mu+\lambda^{\prime}(\eta^{\prime})} &  & \frac{1}{\tau
\mu+\lambda^{\prime}(\eta^{\prime})}\\
-\frac{1}{\tau\mu+\lambda(\eta)} &  & \frac{1}{\tau\mu+\lambda(\eta)} & \\
& \frac{1}{\tau\mu+\lambda^{\prime}(\eta^{\prime})} &  & \frac{1}{\tau
\mu+\lambda^{\prime}(\eta^{\prime})}%
\end{array}
\right)  , \label{VabCondGammaGEN2}%
\end{align}
where%
\begin{equation}
\left\{
\begin{array}
[c]{c}%
\lambda(\eta)=(\omega-g)(1-\tau)+\frac{1-\eta}{\eta}~,\\
\\
\lambda^{\prime}(\eta^{\prime})=(\omega+g^{\prime})(1-\tau)+\frac
{1-\eta^{\prime}}{\eta^{\prime}}~.
\end{array}
\right.  \label{LAMDAS-GEN}%
\end{equation}
Note that Eq.~(\ref{VabCondGammaGEN2}) could have been computed directly from
Eq.~(\ref{totalCM-PR}) by applying the transformations~\cite{NotaLOSS}%
\begin{equation}
\lambda\rightarrow\lambda(\eta),~\lambda^{\prime}\rightarrow\lambda^{\prime
}(\eta^{\prime}). \label{Ltrans}%
\end{equation}

\subsection{Asymptotic generalized key-rate}

From the previous CM we can write the sub-matrix describing Bob's mode%
\begin{equation}
\mathbf{V}_{b|\gamma}=\left(
\begin{array}
[c]{cc}%
\mu-\frac{\tau(\mu^{2}-1)}{2[\tau\mu+\lambda(\eta)]} & \\
& \frac{\tau(\mu^{2}-1)}{2[\tau\mu+\lambda^{\prime}(\eta^{\prime})]}%
\end{array}
\right)  . \label{Vbobgamma}%
\end{equation}
Then, by applying Eq.~(\ref{HET-FORMULA}) to the generalized CM of
Eq.~(\ref{VabCondGammaGEN2}), we derive the doubly-conditional CM of Bob,
conditioned to both relay's and Alice's detections, i.e.,%
\begin{equation}
\mathbf{V}_{b|\gamma\alpha}(\eta,\eta^{\prime})=\left(
\begin{array}
[c]{cc}%
\mu-\frac{\tau(\mu^{2}-1)}{[\tau(\mu+1)+2\lambda(\eta)]} & \\
& \mu-\frac{\tau(\mu^{2}-1)}{[\tau(\mu+1)+2\lambda^{\prime}(\eta^{\prime})]}%
\end{array}
\right)  . \label{Vbgammaalpha}%
\end{equation}

We can now derive the symplectic spectra of the CMs of
Eq.~(\ref{VabCondGammaGEN2}) and~(\ref{Vbgammaalpha}). We find simple
analytical expressions in the limit of large modulation. For $\mathbf{V}%
_{ab|\gamma}$ we have the symplectic eigenvalues $\nu_{1}(\eta)$ and $\nu
_{2}(\eta^{\prime})$, while for $\mathbf{V}_{b|\gamma\alpha}$ we have
$\nu(\eta,\eta^{\prime})$. These eigenvalues can be obtained by applying the
transformations of Eq.~(\ref{Ltrans}) to the Eqs.~(\ref{total-SPECTRUM})
and~(\ref{COND-SPECTRUM}).

Using these spectra, we can compute the corresponding total and conditional
von Neumann entropies and therefore the Holevo bound. In the limit of large
modulation, Eve's Holevo information becomes%
\begin{align}
I_{E|\gamma}(\eta,\eta^{\prime})  &  =\log_{2}\frac{e^{2}\sqrt{\lambda
(\eta)\lambda^{\prime}(\eta^{\prime})}\mu}{4\tau}\nonumber\\
&  -h\left[  \frac{\sqrt{[\tau+2\lambda(\eta)][\tau+2\lambda^{\prime}%
(\eta^{\prime})]}}{\tau}\right]  , \label{HolevoAPP}%
\end{align}
which extends Eq.~(\ref{HOLEVO-BOUND}) to arbitrary efficiencies $\eta$ and
$\eta^{\prime}$.

Similarly, we can extend the formula for Alice and Bob's mutual information,
which here becomes%
\begin{equation}
I_{AB|\gamma}\left(  \eta,\eta^{\prime}\right)  =\log_{2}\frac{\tau\mu}%
{4\sqrt{[\tau+\lambda(\eta)][\tau+\lambda^{\prime}(\eta^{\prime})]}},
\label{IAB-ASY}%
\end{equation}
for large modulation. Combining the previous results, we derive the asymptotic
key-rate in the presence of detector inefficiencies%
\begin{align}
R_{sym}(\eta,\eta^{\prime})  &  =\log_{2}\left[  \frac{\tau^{2}}{e^{2}%
\sqrt{\lambda(\eta)\lambda^{\prime}(\eta^{\prime})[\tau+\lambda(\eta
)][\tau+\lambda^{\prime}(\eta^{\prime})]}}\right] \nonumber\\
&  +h\left[  \frac{\sqrt{[\tau+2\lambda(\eta)][\tau+2\lambda^{\prime}%
(\eta^{\prime})]}}{\tau}\right]  , \label{Rsym-GEN}%
\end{align}
which clearly extends the formula given in Eq.~(\ref{Rsym}).

\subsection{Role of the imperfections on key-rate, security thresholds and
achievable distances}

In this section we study in detail the combined role of the various
experimental limitations and imperfections, confirming\ the main findings
presented in the main body of this paper. We compute the key-rate and the
security thresholds considering not only the realistic quantum efficiency of
the detectors, but also the use of a finite Gaussian modulation and the
non-ideal reconciliation efficiency provided by realistic codes for error
correction and privacy amplification.

\subsubsection{Secret key rate}

In order to extract a secret key, the honest parties must process their
correlated data in stages of perform error correction and privacy
amplification. This data processing is today implemented with a limited
efficiency $\beta\leq1$, for instance $\beta\simeq0.95\div0.97~$%
\cite{leverrier,jouguet}. To include this limitation, we have to multiply
Alice and Bob's mutual information by $\beta$, and consider the realistic
key-rate~\cite{RMP}%
\begin{equation}
R(\beta,\mu,\tau,\omega,g,g^{\prime},\eta,\eta^{\prime})=\beta I_{AB|\gamma
}-I_{E|\gamma}, \label{GEN-KEY-RATE}%
\end{equation}
where $I_{AB|\gamma}$ and $I_{E|\gamma}$ are now computed considering finite
modulation $\mu$ besides non-ideal detector efficiencies $\eta$ and
$\eta^{\prime}$ (clearly these quantities must tend to Eqs.~(\ref{HolevoAPP})
and~(\ref{IAB-ASY}) in the limit of large modulation).

\begin{figure}[t]
\vspace{+0.2cm}
\par
\begin{center}
\includegraphics[width=0.45\textwidth] {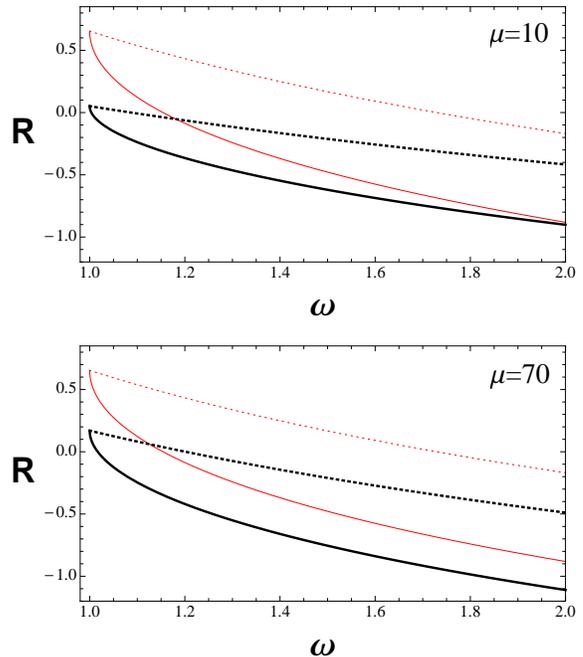}
\end{center}
\par
\vspace{-0.6cm}\caption{(Color online) The key-rate $R$ (bits) is plotted
versus thermal noise $\omega$ for $\mu=10$ (SNU)\textbf{ }(top-panel) and
$\mu=70$ (bottom-panel). Other parameters are $\tau=0.9$, $\beta=0.95$ and
$\eta=\eta^{\prime}=0.98$. We compare two eavesdropping strategies: The
collective one-mode entangling cloner attack $g^{\prime}=g=0$ (dotted black
line) and the two-mode \textquotedblleft negative EPR\textquotedblright%
\ attack $g=-g^{\prime}=-\sqrt{\omega^{2}-1}$ (continuous black line). We see
that the key-rate of the two-mode attack is always lower than that of the
one-mode attack. For comparison, we have also plotted the performances in the
case of ideal reconciliation ($\beta=1$), ideal detectors ($\eta=\eta^{\prime
}=1$) and large modulation ($\mu\gg1$). These ideal performances are
represented by the red curves, dotted for the one-mode attack and continuous
for the two-mode attack.}%
\label{Rate-general}%
\end{figure}In general, the mutual information can be computed from the
formula $I_{AB|\gamma}=\frac{1}{2}\log_{2}\Sigma$, where $\Sigma$ is defined
in Ref.~\cite{RELAY}. Eve's Holevo information can be computed using the
formula of the von Neumann entropy $S=\Sigma_{x}h(x)$, with $h(x)$ defined in
Eq.~(\ref{H}) and applied to the numerical symplectic eigenvalues of the CMs
given in Eqs.~(\ref{VabCondGammaGEN2}) and~(\ref{Vbgammaalpha}). In
Fig.~\ref{Rate-general}, we plot the key-rate of Eq.~(\ref{GEN-KEY-RATE}) as a
function of the thermal noise $\omega$ for two values of the Gaussian
modulation $\mu=10$ (top) and $\mu=70$ (bottom)\textbf{ }vacuum shot noise
unit (SNU), and choosing $\tau=0.9$, $\eta=\eta^{\prime}=0.98$ and
$\beta=0.95$. We see that the key-rate of a negative EPR attack is clearly
lower than that of a collective one-mode attack. This behavior is generic by
varying the previous parameters.

\subsubsection{Security thresholds and achievable distances}

\begin{figure}[t]
\vspace{+0.2cm}
\par
\begin{center}
\includegraphics[width=0.45\textwidth] {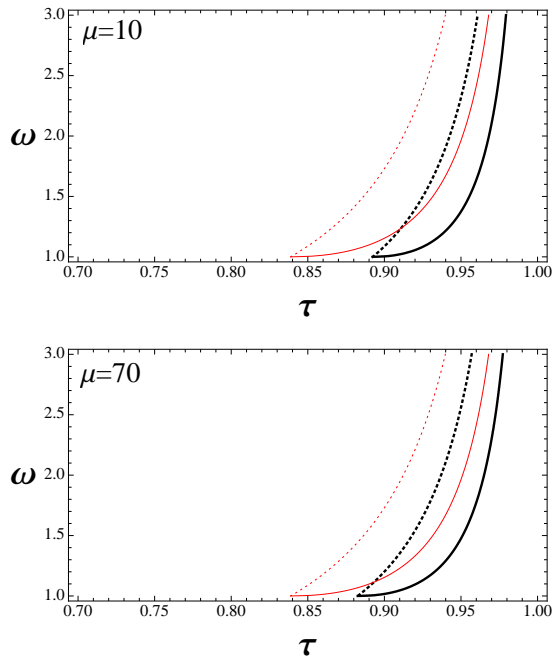}
\end{center}
\par
\vspace{-0.6cm}\caption{(Color online) We plot the security threshold
$\omega=\omega(\tau)$ for $\mu=10$ (top panel) and $\mu=70$ (bottom panel).
Other parameters are $\beta=0.95$ and $\eta=\eta^{\prime}=0.98$. We compare
the two-mode negative EPR attack (continuous black lines) and the one-mode
entangling cloner attack (dotted black lines). Red curves refer to the ideal
case $\beta=\eta=\eta^{\prime}=1$ and $\mu\gg1$.}%
\label{thresholds}%
\end{figure}

Here we study the impact of the experimental limitations on the security
thresholds, comparing the two-mode optimal attack with one-mode collective
attack. The security threshold is obtained by solving the equation%
\begin{equation}
R(\beta,\mu,\tau,\omega,g,g^{\prime},\eta,\eta^{\prime})=0.
\end{equation}
In this equation, we fix the values of the Gaussian modulation ($\mu=10$ or
$70$), the reconciliation efficiency $(\beta=0.95)$, and the quantum
efficiencies $\eta=\eta^{\prime}=0.98$. Then, for each attack, we can write
the security threshold as $\omega=\omega(\tau)$. The comparison is provided in
Fig.~\ref{thresholds}, where we see that the threshold of the optimal two-mode
attack is always lower than the threshold of the one-mode collective attack.

\begin{figure}[t]
\vspace{+0.2cm}
\par
\begin{center}
\includegraphics[width=0.5\textwidth] {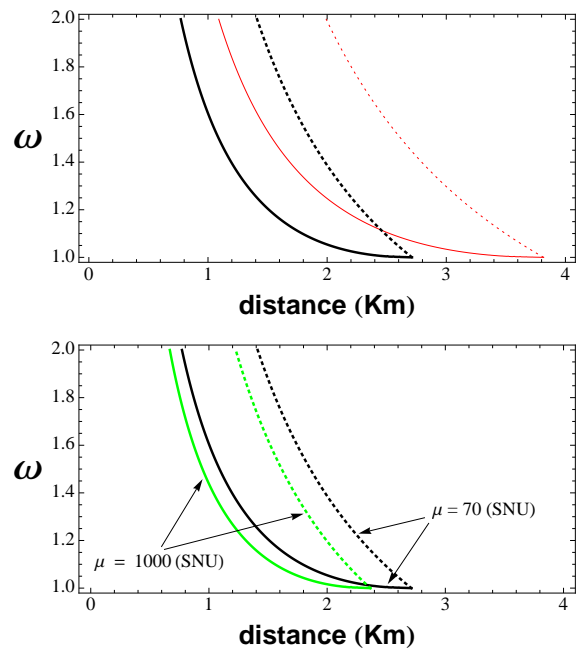}
\end{center}
\par
\vspace{-0.6cm}\caption{(Color online) This figure shows the security
thresholds as $\omega=\omega(d)$, where $d$ is the distance in km, assuming
the loss rate of $0.2$ dB/Km. As in previous figures we compare the two-mode
negative EPR attack (solid lines) with the one-mode entangling-cloner attack
(dotted lines). In the top panel, the red curves refer to the ideal conditions
$\beta=\eta=\eta^{\prime}=1$ and $\mu\gg1$. The black curves refer to the non
ideal case\textbf{ }$\beta=0.95$, $\eta=\eta^{\prime}=0.98$ and $\mu=70$
(SNU)\textbf{. }The bottom panel shows the degradation of the performance as
we increase the modulation from $\mu=70$ (black curves) to $\mu=1000$ (green
curves), for $\beta=0.95$, $\eta=\eta^{\prime}=0.98$.}%
\label{distances}%
\end{figure}

The previous analysis can be performed by expressing the transmissivity in
term of distances. In fact, we may consider $\tau=10^{-\frac{0.2}{10}d}$,
where $d$ is the distance in km, assuming the standard loss rate in fibre of
$0.2$dB/Km. The achievable distances are shown in Fig.~\ref{distances}. We see
that moving from the ideal conditions (red curves, with $\beta=\eta
=\eta^{\prime}=1$ and $\mu\gg1$) the performances deteriorate. All others
curves are obtained for realistic reconciliation efficiencies $\beta=0.95$,
and detectors efficiencies $\eta=\eta^{\prime}=0.98$. The top panel compares
the ideal thresholds with the case $\mu=70$ (black), while in the bottom panel
we show the degradation of the performances while increasing the modulation
from $\mu=70$ (black) to $\mu=1000$ (green). 

It is interesting to note the effect of the reconciliation efficiency on the
optimal modulation variance. For values of $\beta<1$, the optimal modulation
is not infinite. In fact, for the realistic value considered here,
$\beta=0.95$, we have a range of finite modulations $30\lesssim\mu\lesssim70$,
for which the performances are improved.

\subsubsection{Discussion}

Here we summarize some important aspects emerged from this further analysis.
First, we have shown that positive key-rates are still achievable over the
range of metropolitan distances in the presence of various experimental
limitations. Second, the negative-EPR attack, already identified to be the
optimal attack in the ideal case (see main text) continues to be the most
powerful eavesdropping strategy also considering realistic experimental
conditions, i.e., finite modulation, non-ideal reconciliation and non-ideal
detectors. Third, the degradation of the performances of the protocol does not
come from the finite modulation (e.g., we checked that values as low as
$\mu=10$ are still acceptable) but mainly from the quantum efficiencies of the
detectors, $\eta$ and $\eta^{\prime}$, and the reconciliation efficiency
$\beta$ of the classical codes.


\begin{thebibliography}{99}                                                                                               %


\bibitem {ClassCRY}B. Schneier, Applied Cryptography (John Wiley \& Sons, New
York, 1996) p.23.

\bibitem {Gisin}N. Gisin, G. Ribordy, W. Tittel, and H. Zbinden, Rev. Mod.
Phys. \textbf{74}, 145 (2002).

\bibitem {SECOQC}SECOQC, 2007, http://www.secoqc.net

\bibitem {SECOQC2}M. Peev \textit{et al}., New J. Phys. 11, 075001 (2009).

\bibitem {Tokyo1}Tokyo QKD network 2010, www.uqcc.org/QKDnetwork.

\bibitem {endtoend}J. H. Saltzer, D. P. Reed, and D. D. Clark,
\textit{Proceedings of the Second International Conference on Distributed
Computing Systems} (Paris, France, April 8-10, 1981).

\bibitem {Baran}P. Baran, IEEE Trans. on Comm. \textbf{12}, pp. 1-9 (1964).

\bibitem {SidePRL}S. L. Braunstein, and S. Pirandola, Phys. Rev. Lett.
\textbf{108}, 130502 (2012).

\bibitem {EXP1}A. Rubenok, J. A. Slater, P. Chan, I. Lucio-Martinez, and W.
Tittel, Phys. Rev. Lett. \textbf{111}, 130501 (2013).

\bibitem {EXP2}T. Ferreira da Silva, D. Vitoreti, G. B. Xavier, G. C. do
Amaral, G. P. Tempor\~{a}o, and J. P. von der Weid, Phys. Rev. A \textbf{88},
052303 (2013).

\bibitem {RELAY}S. Pirandola, C. Ottaviani, G. Spedalieri, C. Weedbrook, S.L.
Braunstein, S. Lloyd, T. Gehering, C.S. Jacobsen and U.L. Andersen, Nature
Photonics \textbf{9}, 397--402 (2015). See also arXiv.1312.4104.

\bibitem {BraREV2}S. L. Braunstein and P. van Loock, Rev. Mod. Phys.
\textbf{77}, 513 (2005).

\bibitem {RMP}C. Weedbrook, S. Pirandola, R. Garcia-Patron, N. J. Cerf, T. C.
Ralph, J. H. Shapiro, and S. Lloyd, Rev. Mod. Phys. \textbf{84}, 621 (2012).

\bibitem {patron-gaussian}R.~Garc\'{\i}a-Patr\'{o}n and N.J. Cerf, Phys. Rev.
Lett. \textbf{97}, 190503 (2006).

\bibitem {navascues}M. Navascues, F. Grosshans, A.Acin, Phys. Rev. Lett.
\textbf{97}, 190502 (2006).

\bibitem {Li}Z. Li, Y-C. Zhang, F. Xu, X. Peng, and H. Guo, Phys. Rev. A
\textbf{89}, 052301 (2014).

\bibitem {Ma}X-C. Ma, S-H. Sun, M-S. Jiang, M. Gui, and L-M. Liang, Phys. Rev.
A \textbf{89}, 042335 (2014).

\bibitem {NoteEXCESS}Note that Ref.~\cite{Ma} computed the rate of the
protocol at fixed transmissivity $\tau$ and thermal noise $\omega$, since they
fixed both $\tau$ and $\varepsilon=(\omega-1)(1-\tau)/\tau$. However, while
the latter formula provides the excess noise for standard one-way
protocols~\cite{RMP} it does not for the considered relay-based protocol (see
Ref.~\cite{RELAY} for the correct definition of excess noise in this more
complex case).

\bibitem {BellFORMULA}G. Spedalieri, C. Ottaviani, and S. Pirandola, Open
Syst. Inf. Dyn. \textbf{20}, 1350011 (2013).

\bibitem {TwoWay}S. Pirandola, S. Mancini, S. Lloyd, and S. L. Braunstein,
Nature Phys. \textbf{4}, 726 (2008).

\bibitem {Weedbrook2013}C. Weedbrook, C. Ottaviani, and S. Pirandola, Phys.
Rev. A \textbf{89}, 012309 (2014).

\bibitem {Nielsen}M.A. Nielsen and I. L. Chuang, \textit{Quantum Computation
and Quantum Information} (Cambridge University Press, Cambridge, 2000).

\bibitem {Wilde}M. M. Wilde,\textit{\ From Classical to Quantum Shannon
Theory} (Cambridge University Press, Cambridge, 2013).

\bibitem {giedke}G. Giedke and J.I. Cirac, Phys. Rev. A \textbf{66}, 032316 (2002).

\bibitem {Renner}R. Renner, Nature Phys\textit{.} \textbf{3}, 645 (2007).

\bibitem {renner-cirac}R. Renner, J.I. Cirac, Phys. Rev. Lett. 102, 110504 (2009).

\bibitem {canATTACKS}S. Pirandola, S. L. Braunstein, and S. Lloyd, Phys. Rev.
Lett. \textbf{101}, 200504 (2008).

\bibitem {TwomodePRA}S. Pirandola, A. Serafini, and S. Lloyd, Phys. Rev. A
\textbf{79}, 052327 (2009).

\bibitem {NJP2013}S. Pirandola, New J. Phys. \textbf{15}, 113046 (2013).

\bibitem {Weedbrook2010}C.~Weedbrook, S.~Pirandola, S.~Lloyd, and T.C.~Ralph,
Phys. Rev. Lett. \textbf{105}, 110501 (2010).

\bibitem {Weedbrook2012}C.~Weedbrook, S.~Pirandola, and T.C.~Ralph, Phys. Rev.
A \textbf{86}, 022318 (2012).

\bibitem {leverrier}Paul Jouguet, S\'{e}bastien Kunz-Jacques, Anthony
Leverrier, Phys. rev. A \textbf{84, }062317 (2011).

\bibitem {jouguet}P. Jouguet, S. Kunz-Jacques, A. Leverrier, P. Grangier, E.
Diamanti, Nat. Photonics \textbf{7}, 378 (2013).

\bibitem {NotaLOSS}Note that, in the presence of a pure-loss attack
$(\omega=1$ and $g=g^{\prime}=0)$, the non-unit quantum efficiencies of the
detectors can be included in the loss of the channels. For simplicity, for
$\eta=\eta^{\prime}$ we can write
\[
\frac{\tau}{\tau\mu+\lambda(\eta)}=\frac{\tau\eta}{\tau\eta\mu+1-\tau\eta}%
\]
which is equivalent to consider $\tau\eta$ in the CM of Eq.~(\ref{totalCM-PR}).
\end{thebibliography}
\end{document}